\documentclass[11pt]{elsart}

\usepackage{bbm}
\usepackage{soul}
\usepackage{cite}
\usepackage{graphics,color,epsfig,amsmath,amssymb,bm}
\usepackage{feynmp}
\usepackage{subfigure}
\usepackage[english=british]{csquotes}

\newcounter{bla}

\renewcommand{\half}{\ensuremath{\frac{1}{2}}}

\newcommand{\T}{^{T}}

\newcommand{\be}{\begin{equation}}
\newcommand{\ee}{\end{equation}}
\newcommand{\beqna}{\begin{eqnarray}}
\newcommand{\eeqna}{\end{eqnarray}}

\newcommand{\w}{\omega}
\newcommand{\wi}{\omega_i}
\newcommand{\wj}{\omega_j}
\newcommand{\wk}{\omega_k}
\newcommand{\wl}{\omega_l}

\newcommand{\nOverTwo}{\frac{n}{2}}

\allowdisplaybreaks

\begin{document}

\begin{frontmatter}
\hfill{Report Number: Not Applicable} 

\title{A simple expression for the four-point scalar function from Gaussian integrals and Fourier transform}

\author[a]{Kamel Benhaddou}

\address[a]{Unaffiliated \\ almo.cadessi@gmail.com}

\begin{abstract}
Recasting the $N$-point one loop scalar integral from Feynman to Schwinger parameters gives an integrand with a Gaussian form. By application of a Fourier transform, it is easy to derive explicit expressions for the two, three and four-point functions. The Fourier transformation disentangles singularities in the complex plane and extract their contribution as two-point functions in two dimensions. We explicitly derive a one dimensional expression for the (4D) four-point function whose integrand involves only square root and arcsine functions. This report is a condensed version of the approach developed in ~\cite{Benhaddou2016} which does not make use of probabilistic jargon.
\begin{flushleft}
PACS: Not Applicable.
\end{flushleft}


\end{abstract}

\end{frontmatter}

\newpage


\section{Introduction}
There exists a large literature on the problem of $N$-point one loop scalar 
integrals. In \cite{Davydychev1992}, an application of Mellin Barnes lemma allowed a general series representation for the $N$-point function in terms of Lauricella hypergeometric function $F^D_{N}$ or its generalised version. In \cite{SuzukiSantosSchmidt2003}, it is shown that the negative dimension approach, Mellin Barnes representation and Feynman parameters yield the same expression for the $N$-point function. \cite{Fleischer2003} used a recurrence relation between $N$-point functions in  different dimensions to give an explicit representation of the  2-,3-, and 4-point functions in general dimensions $n$. 
Specifically the 4-point function (or box diagram) has received a lot of attention. ~\cite{THOOFT1979,vanOldenborgh1990,DENNER1991,VANOLDENBORGH1992,DennerDittmaier2011} perform integration of the Feynman parameters to produce results in terms of a varying number of polylogarithms with real or complex values of the masses.~\cite{vanOldenborgh1990,VANOLDENBORGH1992} shows how to find expressions that are also numerically stable for computer implementation. More recently, the four-point function with complex internal masses has been revisited in ~\cite{Phan2017ftp,Phan2017}. In \cite{DavydychevDelbourgo1998} the problem is written as a problem of hyperbolic geometry involving the computation of the volume of a simplex in $n$ dimensions. This is obtained when the $N$-point function is cast into the problem of finding the volume of the positive part of an ellipsoid (the volume for which the coordinates are positive).
This has been pursued more recently in ~\cite{Davydychev2017} to reduce the number of variables in term of which results are written.
The same problem can actually  be understood probabilistically and describe the probability that a Gaussian random vector has all its components positive as shown in 
~\cite{Benhaddou2016}. In ~\cite{DavydychevDelbourgo1998}, the geometrical interpretation was pushed as far as $N=4$ but the probabilistic approach in ~\cite{Benhaddou2016}  can easily handle cases of $N>4$ and produce results as integrals of square roots and arcsine functions. In the following we give a general integral formula for the $N$-point scalar function in $n$ dimensions and show how to derive an explicit result for the four-point function in four dimensions.

\section{$N$-point one loop scalar integral}
We  consider the $N$-point one loop scalar integral in 
$n$ dimensions and with each propagator having power $\nu_i$ 
\be
J^{N}(n;\nu_1, \nu_2, \ldots, \nu_N) = \int\,\frac{\mathrm{d}^nq}{i\pi^{\frac{n}{2}}}
\frac{1}{\prod_{i=1}^{N}\left[ (p_i + q)^2 - m_i^2\right]^{\nu_i}}\,
\ee
which after introduction of Feynman parameters takes the following form ~\cite{DavydychevDelbourgo1998}
\begin{align}
\label{JN_feynman}J^{N}(n;\bm{\nu};\Sigma) &= \frac{1}{(-1)^{\nu}} \frac{\Gamma(\nu - n/2)}{\prod_i \Gamma(\nu_i)}
\int_{0}^{1}\mathrm{d}u_1 u_1^{\nu_1-1} \ldots \int_{0}^{1}\mathrm{d}u_N u_N^{\nu_N-1}   
\frac{\delta\left( \sum\limits_{i=1}^{N} u_i -1\right)}{\left( -\sum\limits_{j<\ell}u_j u_{\ell} k^2_{j\ell} + \sum\limits_i u_i m_i^2   \right)^{\nu - \frac{n}{2}}}\,, \\
\label{JN_matrix}&= (-1)^{-\nu} \frac{\Gamma(\nu - n/2)}{\prod_i \Gamma(\nu_i)}
\int_{0}^{1}\mathrm{d}u_1 u_1^{\nu_1-1} \ldots \int_{0}^{1}\mathrm{d}u_N u_N^{\nu_N-1}   
\frac{\delta\left( \sum\limits_{i=1}^{N} u_i -1\right)}{\left(u\T . \Sigma .u \right)^{\nu - \frac{n}{2}}}\,,
\end{align}
where $(\sum\limits_i \nu_i =\nu$) and the matrix $\Sigma$ has components
\begin{align}
\Sigma_{ii} &= m_i^2\,, \nonumber \\
\Sigma_{j\ell} &= m_j m_\ell c_{j\ell}   & j\neq \ell\,, \nonumber \\
c_{j\ell} &= \frac{m_j^2 + m_{\ell}^2 - k_{j\ell}^2}{2 m_j m_{\ell}}\,, \nonumber \\
k_{j\ell}^2 &= \left( p_j - p_{\ell}\right)^2\,. \nonumber
\end{align}
The $c_{j\ell}$ can be understood as cosines of some angles $\tau_{j\ell}$ ~\cite{DavydychevDelbourgo1998}
\be\label{def_costau}
c_{j\ell} = \cos\tau_{j\ell} 
= \left\{ \begin{array}{c} \;\; 1, \;\;\; k_{j\ell}^2=(m_j-m_{\ell})^2 \\
                               -1, \;\;\; k_{j\ell}^2=(m_j+m_{\ell})^2 
          \end{array} \right. \; .
\ee
The corresponding angles $\tau_{j\ell}$ are ~\cite{DavydychevDelbourgo1998}
\be\label{def_tau}
\tau_{j\ell}= \arccos(c_{j\ell}) 
=\arccos\left(\frac{m_j^2+m_{\ell}^2-k_{jl}^2}{2m_j m_{\ell}}\right)
= \left\{ \begin{array}{c}   0, \;\;\; k_{jl}^2=(m_j-m_{\ell})^2 \\
                           \pi, \;\;\; k_{jl}^2=(m_j+m_{\ell})^2
          \end{array} \right. \; .
\ee
The angles $\tau_{j\ell}$ can be analytically continued when the $c_{j\ell}$ are not 
in the range $[-1,1]$. When $k_{j\ell}^2<(m_j-m_l)^2$,  the $c_{jl}$ are greater than 1 and the angles $\tau_{j\ell}$ are given by ~\cite{DavydychevDelbourgo1998}
\begin{align}
\tau_{j\ell} &=-\mathrm{i}\; \mathrm{Arch}(c_{j\ell})\,,\\
&= -\frac{\mathrm{i}}{2} 
\ln\left( 
\frac{m_j^2+m_{\ell}^2-k_{j\ell}^2+\sqrt{\lambda(m_j^2,m_{\ell}^2,k_{j\ell}^2)}}
     {m_j^2+m_{\ell}^2-k_{j\ell}^2-\sqrt{\lambda(m_j^2,m_{\ell}^2,k_{j\ell}^2)}}
\right) ,
\end{align}
where $\lambda(x,y,z)$ is the K\"{a}llen function
\be
\lambda(x,y,z) = x^2 + y^2 + z^2 -2xy -2xz -2yz\,.
\ee
When $k_{j\ell}^2>(m_j+m_{\ell})^2$, the $c_{j\ell}$ are smaller than -1 and the angles $\tau_{j\ell}$ are given by ~\cite{DavydychevDelbourgo1998}
\begin{align}
\tau_{j\ell} &=\pi+\mathrm{i}\; \mathrm{Arch}(-c_{j\ell})\,, \\
&= \pi + \frac{\mathrm{i}}{2}
\ln\left(
\frac{k_{jl}^2-m_j^2-m_{\ell}^2+\sqrt{\lambda(m_j^2,m_{\ell}^2,k_{j\ell}^2)}}
     {k_{jl}^2-m_j^2-m_{\ell}^2-\sqrt{\lambda(m_j^2,m_{\ell}^2,k_{j\ell}^2)}}
\right)\,.
\end{align}
When the angles  $\tau_{j\ell}$ are real (as in Eq.(\ref{def_tau})), the $c_{j\ell}$ are in the range $[-1,1]$ and the matrix $\Sigma$ in Eq.(\ref{JN_matrix}) can be interpreted as a covariance matrix ~\cite{Benhaddou2016}. Let us consider a random vector $Z$ of size $N$, whose components are normal random variables $Z_i$ with 0 mean and variance $\sigma^2_i = m_i^2>0$. When the $c_{j\ell}$ are in the range $[-1,1]$, we can interpret the $c_{j\ell}$ as the correlation between $Z_j$ and $Z_{\ell}$. When the $c_{j\ell}$ are complex numbers, the angles  $\tau_{j\ell}$ are given by the arccosine function in the complex plane whose logarithmic (principal values) form is as follows ~\cite[(Eq.)4.23.22-25]{NISTDLMF}
\begin{align}
\arccos(z) &= \half\pi + i \ln\left(i z + (1-z^2)^{\half}\right)\,, & z\in\mathbb{C}\setminus\, ]-\infty,-1[\, \cup\, ]1,\infty[\,, \nonumber \\
&= -2i \ln\left(\left(\frac{1+z}{2}\right)^\half + i \left(\frac{1-z}{2}\right)^\half \right)\,, & z\in\mathbb{C}\setminus\, ]-\infty,-1[\, \cup\, ]1,\infty[\,, \nonumber \\
&= \mp i \ln\left(x + (x^2-1)^{\half}\right)\,, & x\in\,[1,\infty[\,, \nonumber \\
&= \pi \mp i \ln\left(-x + (x^2-1)^{\half}\right)\,, & x\in\,]-\infty, -1]\,. 
\end{align}
When $z=x+iy$ and $z \notin\, ]1,\infty[$, we can compute the real and imaginary part of $\arccos(z)$
\begin{align}
\arccos(z) &= \arccos(\beta) - i \mathrm{sign}(y)\ln\left(\alpha + (\alpha^2-1)^\half\right)\,,\nonumber \\
\alpha     &= \half\left( (x+1)^2 +y^2\right)^\half + \half\left( (x-1)^2 +y^2\right)^\half\,, \nonumber \\
\beta      &= \half\left( (x+1)^2 +y^2\right)^\half - \half\left( (x-1)^2 +y^2\right)^\half\,.
\end{align}
Using the relation
\be\label{Schwinger_parameter}
\frac{\Gamma(\alpha)}{A^\alpha} = \int_{0}^{\infty}\mathrm{d} \tau\, \tau^{\alpha-1} e^{-\tau A}\,,
\ee
Eq.(\ref{JN_matrix}) is rewritten as
\be
J^{N}(n;\bm{\nu};\Sigma) =  \frac{(-1)^{-\nu}}{\prod_i \Gamma(\nu_i)}
\int_{0}^{1}\mathrm{d}u_1 u_1^{\nu_1-1} \ldots \int_{0}^{1}\mathrm{d}u_N u_N^{\nu_N-1}   
\delta\left( \sum\limits_{i=1}^{N} u_i -1\right)
\int_{0}^{\infty}\, \mathrm{d}\tau\,  \tau^{\nu -\frac{n}{2}-1} e^{-\tau u\T .\Sigma. u }\,.
\ee
We change the variable $\tau$ to $v$ such that $\tau = \frac{v^2}{2}$ and obtain
\be\label{J_N_v_integration}
J^{N}(n;\bm{\nu};\Sigma) = (-1)^{-\nu}
\int\displaylimits_{0}^{1}\prod_{i=1}^{N}\mathrm{d}u_i u_i^{\nu_i-1}
\frac{\delta(\sum\limits_{i=1}^{N} u_i -1)}
{2^{\nu -\nOverTwo-1}\prod_i \Gamma(\nu_i)} 
\int\displaylimits_{0}^{\infty}\, \mathrm{d}v\, v^{2\nu -n-1} e^{-\half (vu)\T .\Sigma. (vu) }\,.
\ee
We make the change of variable $x_i = v u_i$ and get
\beqna
J^{N}(n;\bm{\nu};\Sigma) &=& \frac{(-1)^{-\nu}}{2^{\nu -\nOverTwo-1}\prod_i \Gamma(\nu_i)}
\int_{0}^{\infty}\prod_{i=1}^{N}\mathrm{d}x_i x_i^{\nu_i-1}
\int\displaylimits_{0}^{\infty}\, \mathrm{d}v\, v^{\nu-n-1} \delta\left( \frac{1}{v}\sum\limits_{i=1}^{N} x_i -1\right) e^{-\half x\T .\Sigma. x }\,.
\eeqna
We can rewrite the delta function as
\be
\delta\left( \frac{1}{v}\sum\limits_{i=1}^{N} x_i -1\right) =
v \delta\left(\sum\limits_{i=1}^{N} x_i -v\right)\,,
\ee
which forces us to set $v =\sum\limits_{i=1}^{N} x_i$ in the integrand. We have
\be\label{JN_proba}
J^{N}(n;\bm{\nu};\Sigma) =
(-1)^{-\nu}\mathcal{N}
\int_{-\infty}^{\infty}\prod_{i=1}^{N}\mathrm{d}x_i\, 
\frac{\prod_{i=1}^{N}\left(x_i^{\nu_i-1} \, \mathbbm{1}_{\{x_i>0\}}\right)}{2^{\nu -\nOverTwo-1}\prod_i \Gamma(\nu_i)} \left( \sum\limits_{i=1}^{N} x_i\right)^{\nu -n}
\, \frac{e^{-\half x\T .R^{-1}.x}}{\mathcal{N}}\,,
\ee
with $R = \Sigma^{-1}$ and $\mathcal{N} = (2\pi)^{\frac{N}{2}}|R|^{\half}$. The exponential in the integrand is the probability density of the multivariate normal distribution also known as multivariate Gaussian function.

\section{Explicit evaluation of $N$-point functions}
To find explicit expressions we follow the method of ~\cite{Childs1967} where Fourier transform methods are used. Recall Eq.(\ref{JN_proba}) 	
\be
J^{N}(n;\bm{\nu};\Sigma)=
\frac{(-1)^{-\nu}}{2^{\nu -\nOverTwo-1}\prod_i\Gamma(\nu_i)}
\frac{(2\pi)^{\frac{N}{2}}}{|\Sigma|^{\half}} 
\int_{-\infty}^{\infty}\prod\limits_{i=1}^{N}\mathrm{d} z_i\,
\prod\limits_{i=1}^{N}U(z_i)\,
\frac{\prod_i z_{i}^{\nu_i-1}\left(\sum_i z_i\right)^{\nu-n}}{(2\pi)^{\frac{N}{2}}|R|^{\half}}e^{-\half z^{T}.R^{-1}. z}\,,
\ee
where $U(z_i) = \mathbbm{1}_{z_i>0}$. The $N$ dimensional Fourier transform $\hat{f}(w)$ of a function $f(x)$ is defined as
\be
\hat{f}(w) = \int_{-\infty}^{\infty}\mathrm{d}^Nx\, f(x) e^{-iw\T .x}\,,
\ee 
and its inverse is
\be
f(x) = \frac{1}{(2\pi)^N}\int_{-\infty}^{\infty}\mathrm{d}^N w\,  \hat{f}(w) e^{+iw\T .x}\,.
\ee 
both Fourier transform of the unit step(unidimensional) and Gaussian functions (multidimensional)  are known ~\cite{Hansen2014,Tilman2015}
\begin{align}
\hat{U}(w) &= \frac{1}{i} \frac{1}{\w - i\epsilon}\,, \nonumber \\
           &= \frac{1}{i}  \left( i \pi \delta(w) +\mathrm{PV}\frac{1}{w} \right)\,, \nonumber \\
					 &=  \pi \delta(w) +\mathrm{PV}\left(\frac{1}{i w}\right) \,, \\
					 &=  \pi \delta(w) -i \lim_{\epsilon \to 0}\frac{w}{w^2 + \epsilon^2}\,,\nonumber\\
\mathcal{F}\left(
{\frac{1}{(2\pi)^{\frac{N}{2}}|R|^{\half}}e^{-\half z\T .R^{-1}. z}}
\right)(w) &=  e^{-\half w\T .R. w}\,.
\end{align}
In the following, the limit ($\epsilon \to 0$) is assumed even if not written explicitly. Moreover the Fourier transform satisfies
\be
\mathcal{F}\left[x_k f(x) \right](w) = i\frac{\mathrm{\partial}}{\partial w_k}\hat{f}(w)\,.
\ee 
The Parseval relation states
\be
\int_{\infty}^{\infty}\mathrm{d}^Nx\,  f(x) g(x) = 
\frac{1}{(2\pi)^N}\int_{\infty}^{\infty}\mathrm{d}^N w\,  \hat{f}(w) \hat{g}(-w)\,,  
\ee
so we get
\beqna
J^{N}(n;\bm{\nu};\Sigma)&=&
\frac{(-1)^{-\nu}}{2^{\nu -\nOverTwo-1}\prod_i\Gamma(\nu_i)}
\frac{(2\pi)^{\frac{N}{2}}}{|\Sigma|^{\half}} 
\frac{1}{(2\pi)^{N}}\int_{-\infty}^{\infty}\prod\limits_{j=1}^{N}\mathrm{d} w_j\,
\left[
\prod\limits_{j=1}^{N}\left( \pi \delta(\omega_j) - \frac{i}{\omega_j} \right)
\right. \nonumber \\
&& \label{N-point_fourier} \left.
\prod\limits_{j=1}^{N} \left(i\frac{\partial }{\partial \omega_j}\right)^{\nu_j-1}
\left(i\sum\limits_{j=1}^{N}\frac{\partial }{\partial \omega_j}\right)^{\nu-n}
\right]
e^{-\half w^{T}\Sigma^{-1} w}\,.
\eeqna

\subsection{Two-point function in dimension $n=2$}
The Fourier integral we need to compute is (with $R = \Sigma^{-1}$)
\be
F_2 = \frac{1}{(2\pi)^{2}}\int_{-\infty}^{\infty} \mathrm{d} w_1\mathrm{d} w_2\,
\left( \pi \delta(w_1) +\frac{1}{i w_1} \right)
\left( \pi \delta(w_2) +\frac{1}{i w_2} \right)
e^{-\half \left( R_{11}w_1^2 + 2 R_{12}w_1w_2 + R_{22} w_2^2 \right)}\,.
\ee
The matrix $R$ has components
\begin{align}
R_{11} &= \frac{ m_2^2}{\Delta^{(2)}}\,,\nonumber \\
R_{22} &=  \frac{m_1^2}{\Delta^{(2)}}\,,  \nonumber \\
R_{12} &= - \frac{m_1 m_2 c_{12}}{\Delta^{(2)}}\,,  \nonumber \\
\Delta^{(2)} &= m_1^2 m_2^2 \left( 1- c_{12}^2\right) =  m_1^2 m_2^2 \sin^2 \tau_{12}\,. \nonumber 
\end{align}
Two terms are null because they involve an odd power of $w_i$ and the product of the two delta functions gives a constant
\be
F_{2} = \frac{1}{4}  -\frac{1}{4\pi^2}\int_{-\infty}^{\infty} \mathrm{d} w_1\mathrm{d} w_2\,
\frac{e^{-\half \left( R_{11}w_1^2 + 2 R_{12}w_1w_2 + R_{22} w_2^2 \right)}}{w_1 w_2}\,.
\ee
We have ~\cite[reproduced here in the appendix]{Childs1967}
\be
\int_{-\infty}^{\infty} \mathrm{d} w_1\mathrm{d} w_2\,
\frac{e^{-\half \left( R_{11}w_1^2 + 2 R_{12}w_1w_2 + R_{22} w_2^2 \right)}}{w_1 w_2}
= -2\pi \arcsin\left( \frac{R_{12}}{\sqrt{R_{11}R_{22}}}\right)\,.
\ee
The two-point function is 
\begin{align}
J^{2}(2;\bm{1};\Sigma) &=
\frac{2\pi}{m_1 m_2 \sin \tau_{12}}
\left( \frac{1}{4} +\frac{1}{2\pi}\arcsin\left(\frac{R_{12}}{\sqrt{R_{11}R_{22}}}\right)\right)\,,\\
&=\frac{\tau_{12}}{m_1 m_2 \sin \tau_{12}}\,,
\end{align}
where we have used
\be\label{arcsin}
\arcsin(z) = -\frac{\pi}{2} + \arccos(-z)\,,
\ee
to get the last line, which is exactly the same expression as Eq.(4.3) derived in ~\cite{DavydychevDelbourgo1998} (modulo an $i\pi^{\frac{n}{2}}$ factor that we included in our definition of $J^{N}(n, \bm{\nu}, \Sigma)$).

\subsection{Three-point function in dimension $n=3$}
The matrix $R$ has components
\begin{align}
R_{11} &= m_2^2m_3^2(1 - c_{23}^2)/ \Delta^{(3)}\,,  \nonumber \\
R_{22} &= m_1^2m_3^2(1 - c_{13}^2)/ \Delta^{(3)}\,,  \nonumber \\
R_{33} &= m_1^2m_2^2(1 - c_{12}^2)/ \Delta^{(3)}\,,  \nonumber \\
R_{12} &= m_1 m_2 m_3^2 \left( c_{13}c_{23} - c_{12}\right)/ \Delta^{(3)}\,, \nonumber \\
R_{13} &= m_1 m_2^2 m_3 \left( c_{12}c_{23} - c_{13}\right)/ \Delta^{(3)}\,, \nonumber \\
R_{23} &= m_1^2 m_2 m_3 \left( c_{12}c_{13} - c_{23}\right)/ \Delta^{(3)}\,, \nonumber \\
\Delta^{(3)}&= m_1^1 m_2^2 m_3^2 \left( 1 - c_{12}^2 - c_{13}^2 - c_{23}^2 +2c_{12}c_{13}c_{23}\right)\,,\nonumber \\
\Delta^{(3)}&= m_1^1 m_2^2 m_3^2 \, D^{(3)}\,.\nonumber 
\end{align}
In Eq.(\ref{N-point_fourier}), 4 terms are null and we get a constant and 3 terms that are exactly like the 2 point  integral
\begin{align}
J^{3}(3;\bm{1};\Sigma) &= \frac{(-1)(2\pi)^{\frac{3}{2}}}{2^{\half}m_1m_2m_3 \sqrt{D^{(3)}}}
\left[ \frac{1}{8} 
+ \frac{1}{4\pi}\arcsin\left(\frac{R_{12}}{\sqrt{R_{11}R_{22}}}\right) \right.\nonumber \\ 
&+ \left.\frac{1}{4\pi}\arcsin\left(\frac{R_{13}}{\sqrt{R_{11}R_{33}}}\right)
+ \frac{1}{4\pi}\arcsin\left(\frac{R_{23}}{\sqrt{R_{22}R_{33}}}\right) 
\right]\,.
\end{align}
Using the relation Eq.(\ref{arcsin}) and the notation in ~\cite{DavydychevDelbourgo1998}
\begin{align}
\cos \Psi_{12}	&=	\frac{R_{12}}{\sqrt{R_{11}R_{22}}} = \frac{c_{12} - c_{13}c_{23}}{\sin\tau_{13} \sin\tau_{23}}\,,\\
\cos\Psi_{13}		&=	\frac{R_{13}}{\sqrt{R_{11}R_{33}}} = \frac{c_{13} - c_{12}c_{23}}{\sin\tau_{12} \sin\tau_{23}}\,,\\
\cos\Psi_{23}		&= 	\frac{R_{23}}{\sqrt{R_{22}R_{33}}} = \frac{c_{23} - c_{12}c_{13}}{\sin\tau_{13} \sin\tau_{12}}\,,\\
\Omega^{(3)}  	&= 	\Psi_{12}+ \Psi_{13}  + \Psi_{23}  -\pi\,,
\end{align}
we get  
\be
J^{3}(3;\bm{1};\Sigma) = -\frac{\pi^{\half}}{2 m_1 m_2 m_3} \frac{\Omega^{(3)}}{\sqrt{\Delta^{(3)}}}\,,
\ee
exactly as Eq.(5.6) in ~\cite{DavydychevDelbourgo1998}.

\subsection{Three-point function in 2$D$}
In this case in Eq.(\ref{JN_proba}), the term $(\sum\limits_{i=1}^{3}x_i)^{\nu -n}$ contributes to the integral. 
\be
J^{3}(2;\bm{1};\Sigma) = \frac{(-1)(2\pi)^{\frac{3}{2}}}{2m_1m_2m_3 \sqrt{D^{(3)}}}
\frac{1}{(2\pi)^3} \int_{-\infty}^{\infty}\mathrm{d}^3\omega 
\prod\limits_{j=1}^{3}\left[ \pi \delta(i \omega_j) +\frac{1}{\omega_j}\right]
\left( \sum\limits_{j=1}^{3} i\frac{\partial}{\partial\omega_j}\right)e^{\omega^{T}R\omega}\,.
\ee
We expand and keep the non zero terms
\begin{align}
J^{3}(2;\bm{1};\Sigma) &= \frac{(-1)(2\pi)^{\frac{3}{2}}}{2m_1m_2m_3 \sqrt{D^{(3)}}}
\frac{1}{(2\pi)^3}
\left[ 
 (R_{11}+R_{12}+R_{13})( -\pi^2 I_{1} + I_{23}^{1})  \right. \nonumber \\ 
&+  (R_{12}+R_{22}+R_{23})( -\pi^2 I_{2} + I_{13}^{2}) \nonumber \\ 
&+ \left. (R_{13}+R_{23}+R_{33})(- \pi^2 I_{3} + I_{12}^{3} \right]\,,  \nonumber
\end{align}
with
\begin{align}
I_{j} &= 
\frac{\sqrt{2\pi}}{\sqrt{R_{jj}}}\,,\nonumber \\
I_{ij}^{k} &= 
 -\frac{(2\pi)^{\frac{3}{2}}}{\sqrt{R_{kk}}}
\arcsin\left( \frac{\rho_{ij} - \rho_{ik}\rho_{jk}}{\sqrt{1-\rho_{ik}^2}\sqrt{1-\rho_{jk}^2}}\right)\,,\nonumber \\ 
\rho_{ij} &= \frac{R_{ij}}{\sqrt{R_{ii}R_{jj}}}\,. \nonumber
\end{align}
Explicitly
\begin{align}\label{eq:3-point-2D}
J^{3}(2;\bm{1};\Sigma) &= \frac{(2\pi)^{\frac{3}{2}}}{2^{\half}m_1m_2m_3 \sqrt{D^{(3)}}}
\left[ 
 \frac{(R_{11}+R_{12}+R_{13})}{\sqrt{\pi R_{11}}}
\left(\frac{1}{8} + \frac{1}{4\pi} \arcsin\left( \frac{\rho_{23} - \rho_{12}\rho_{13}}{\sqrt{1-\rho_{13}^2}\sqrt{1-\rho_{23}^2}}\right) \right)  \right. \nonumber \\ 
&+  \frac{(R_{12}+R_{22}+R_{23})}{\sqrt{\pi R_{22}}}
\left(\frac{1}{8} + \frac{1}{4\pi} \arcsin\left( \frac{\rho_{13} - \rho_{12}\rho_{23}}{\sqrt{1-\rho_{12}^2}\sqrt{1-\rho_{23}^2}}\right) \right)   \nonumber \\ 
&+ \left. \frac{(R_{13}+R_{23}+R_{33})}{\sqrt{\pi R_{33}}}
\left(\frac{1}{8} + \frac{1}{4\pi} \arcsin\left( \frac{\rho_{12} - \rho_{13}\rho_{23}}{\sqrt{1-\rho_{13}^2}\sqrt{1-\rho_{23}^2}}\right) \right)  \right]\,. 
\end{align}
In ~\cite[section 4.2]{Davydychev2006},  the 3-point function in 2$D$ is also written as a sum of three terms. Each term looks like a 2-point function in 2$D$. The 3-point function in 3$D$ is a linear combination of 2$D$ 2-point functions with coefficients written using the $\tau_{j\ell}$ variables. Eq.(\ref{eq:3-point-2D}) looks very similar, but written explicitly in terms of the $c_{j\ell}$ variables which are directly linked to the kinematical invariants.~\cite{Davydychev2006} also presents results for the 3-point function in dimension $n=4,5$.

\subsection{Four-point function in dimension $n=4$}
We write
\be
|\Sigma|^{\half} = m_1 m_2 m_3 m_4 \sqrt{D^{(4)}}\,.
\ee
In Eq.(\ref{N-point_fourier}), 8 terms are null and we get a constant and 6 terms that are exactly like the 2 point  integral plus a integral that involves all the variables $\wi$.
we have
\begin{align}
J^{4}(4;\bm{1};\Sigma) &= \frac{2\pi^2}{m_1 m_2 m_3 m_4 \sqrt{D^{(4)}}}
\, \frac{1}{16} \, \left[ 1 
+ \frac{2}{\pi}\arcsin\left(\frac{R_{12}}{\sqrt{R_{11}R_{22}}}\right) \right.\nonumber \\ 
&+ \left.\frac{2}{\pi}\arcsin\left(\frac{R_{13}}{\sqrt{R_{11}R_{33}}}\right)
+ \frac{2}{\pi}\arcsin\left(\frac{R_{14}}{\sqrt{R_{11}R_{44}}}\right)  \right. \nonumber \\
&+ \left.\frac{2}{\pi}\arcsin\left(\frac{R_{23}}{\sqrt{R_{22}R_{33}}}\right)
+ \frac{2}{\pi}\arcsin\left(\frac{R_{24}}{\sqrt{R_{22}R_{44}}}\right) \right. \nonumber \\
&+ \left. \frac{2}{\pi}\arcsin\left(\frac{R_{34}}{\sqrt{R_{33}R_{44}}}\right)  
+ \frac{1}{\pi^4}I_{1234}
\right]\,.
\end{align}
In the appendix it is shown that $I_{1234}$ is the sum of 3 integrals
\begin{align}
\frac{1}{\pi^4} I_{1234} &= \frac{4\rho_{12}}{\pi^2}\int_{0}^{1}\frac{\mathrm{d}u}{\sqrt{1-\rho_{12}^2u^2}}
\,\arcsin\left( \frac{\alpha_{34}(u)}{\sqrt{\alpha_{33}(u)\alpha_{44}(u)}}\right)\nonumber \\
&+
\frac{4\rho_{13}}{\pi^2}\int_{0}^{1}\frac{\mathrm{d}u}{\sqrt{1-\rho_{13}^2u^2}}
\,\arcsin\left( \frac{\beta_{24}(u)}{\sqrt{\beta_{22}(u)\beta_{24}(u)}}\right) \nonumber \\
&+
\frac{4\rho_{14}}{\pi^2}\int_{0}^{1}\frac{\mathrm{d}u}{\sqrt{1-\rho_{14}^2u^2}}
\,\arcsin\left( \frac{\gamma_{23}(u)}{\sqrt{\gamma_{22}(u)\gamma_{33}(u)}}\right)\,,
\end{align}
with
\begin{align}
\alpha_{33} &= (1- \rho_{23}^2) + u^2(2\rho_{12}\rho_{13}\rho_{23} - \rho_{12}^2 - \rho_{13}^2) \,, \nonumber \\
\alpha_{34} &= (\rho_{34}- \rho_{23}\rho_{24}) + u^2(\rho_{12}\rho_{13}\rho_{24} + \rho_{12}\rho_{14}\rho_{23} - \rho_{13}\rho_{14} - \rho_{12}^2\rho_{34}) \,, \nonumber \\
\alpha_{44} &= (1- \rho_{24}^2) + u^2(2\rho_{12}\rho_{14}\rho_{24} - \rho_{12}^2 - \rho_{14}^2) \,, \nonumber
\end{align}
\begin{align}
\beta_{22} &= (1- \rho_{23}^2) + u^2(2\rho_{13}\rho_{12}\rho_{23} - \rho_{13}^2 - \rho_{12}^2) \,, \nonumber \\
\beta_{24} &= (\rho_{24}- \rho_{23}\rho_{34}) + u^2(\rho_{13}\rho_{12}\rho_{34} + \rho_{13}\rho_{14}\rho_{23} - \rho_{12}\rho_{14} - \rho_{13}^2\rho_{24}) \,, \nonumber \\
\beta_{44} &= (1- \rho_{34}^2) + u^2(2\rho_{13}\rho_{14}\rho_{34} - \rho_{13}^2 - \rho_{14}^2) \,, \nonumber
\end{align}
\begin{align}
\gamma_{22} &= (1- \rho_{24}^2) + u^2(2\rho_{14}\rho_{12}\rho_{24} - \rho_{14}^2 - \rho_{12}^2) \,, \nonumber \\
\gamma_{23} &= (\rho_{23}- \rho_{24}\rho_{34}) + u^2(\rho_{14}\rho_{12}\rho_{34} + \rho_{14}\rho_{13}\rho_{24} - \rho_{12}\rho_{13} - \rho_{14}^2\rho_{23}) \,, \nonumber \\
\gamma_{33} &= (1- \rho_{34}^2) + u^2(2\rho_{14}\rho_{13}\rho_{34} - \rho_{14}^2 - \rho_{13}^2) \,, \nonumber
\end{align}
where
\begin{align}
\rho_{ij} &= \frac{R_{ij}}{\sqrt{R_{ii}R_{jj}}}\,,\nonumber\\
&= \frac{\Delta_{ij}}{\sqrt{\Delta_{ii}\Delta_{jj}}}\,.
\end{align}
$\Delta_{ij}$ is the $(i,j)$ cofactor of the matrix $\Sigma$. This formula is not explicitly symmetric with respect to the indices $(1,2,3,4)$ because we have chosen to transform the denominator $\omega_1$. The final result represents three integrals of arcsine and square root functions whose values in the complex plane are known. In ~\cite{DavydychevDelbourgo1998}, the geometrical interpretation worked easily for the 2- and 3-point functions. For the 4-point function in dimension 4, the computation of the volume of a four dimensional tetrahedron is quite complicated. Using Fourier transforms, allows us to circumvent difficult geometrical decomposition. For $N\geq 5$, the geometrical interpretations seems to become unrealisable, while Fourier transform still performs well ~\cite{Benhaddou2016}. The Fourier approach distangles the singularity in the Fourier space to produce contributions to the 4-point function that can be recognised as 2-point functions in dimension $n=2$. In ~\cite{Davydychev2017}, splitting of the simplex allowed writing the final result in term of hypergeometriuc functions of 3 variables for the 4-point function in dimension $n$. In our case, specialising to $n=4$, our final result is a sum of univariate functions and one dimensional integrals. We recall the behaviour of the arcsine function (principal values) in the complex plane ~\cite[(Eq.)4.23.19-21]{NISTDLMF}
\begin{align}
\arcsin(z) &= - i \ln\left(i z + (1-z^2)^{\half}\right)\,, & z\in\mathbb{C}\setminus\, ]-\infty,-1[\, \cup\, ]1,\infty[\,, \nonumber \\
&= \half\pi \pm i \ln\left(x + (x^2-1)^{\half}\right)\,, & x\in\,[1,\infty[\,, \nonumber \\
&= -\half\pi \pm i \ln\left(-x + (x^2-1)^{\half}\right)\,, & x\in\,]-\infty, -1]\,. 
\end{align}
When $z=x+iy$ and $z \notin\, ]1,\infty[$, we can compute the real and imaginary part of $\arcsin(z)$ ~\cite[(Eq.)4.23.34]{NISTDLMF}
\begin{align}
\arcsin(z) &= \arcsin(\beta) + i \mathrm{sign}(y)\ln\left(\alpha + (\alpha^2-1)^\half\right)\,,\nonumber \\
\alpha     &= \half\left( (x+1)^2 +y^2\right)^\half + \half\left( (x-1)^2 +y^2\right)^\half\,, \nonumber \\
\beta      &= \half\left( (x+1)^2 +y^2\right)^\half - \half\left( (x-1)^2 +y^2\right)^\half\,.
\end{align}

\section{Conclusion}
Recasting the $N$-point function into a Gaussian integral and using Fourier transforms, it was possible to compute $N$-point functions in integer dimension.
Explicit formulae were presented for $N=2,3,4$. The main achievement is the simple expression derived for the four-point function. The final result is a sum of arsine functions and one dimensional integrals involving square root and arcsine functions. The analytical continuation of the arcsine function in well known in the complex plane. We hope to write a program to implement the result on a computer. Following this approach ~\cite{Benhaddou2016} it is possible to get $\epsilon$ expansions as well as computing tensor integrals.

\section*{Acknowledgements}
I would like to thank Mama Benhaddou for encouragements. 

\renewcommand \thesection{\Alph{section}}
\renewcommand{\theequation}{\Alph{section}.\arabic{equation}}
\setcounter{section}{0}
\setcounter{equation}{0}

\section{Computation of $I_{ij}$}
We define
\be
\rho_{ij} = \frac{R_{ij}}{\sqrt{R_{ii}}\sqrt{R_{jj}}}\,.
\ee
$I_{ij}(R)$ is given by
\be
I_{ij}(R) =\int_{-\infty}^{\infty}\mathrm{d}^2\w
\frac{1}{\w_i \w_j }  \, e^{-\half \sum\limits_{m,n} \w_m R_{mn} \w_n}\,.
\ee
We change variable $w_m \rightarrow \frac{w_m}{\sqrt{R_{mm}}}$ 
\be
I_{ij}(R) =
\int_{-\infty}^{\infty}\mathrm{d}^2\w
\frac{1}{\w_i \w_j }  \, e^{-\half \left(\w_i^2 + \w_j^2  + 2 \rho_{ij} \w_i \w_j \right)}\,.
\ee
For a function $f(\rho)$ we can write
\be
f(\rho) = f(0)  + \int_{0}^{\rho}f'(u)\mathrm{d}u\,,
\ee
where in our case the function $f(\rho)$ is $I_{ij}$ as a function 
of $\rho_{ij}$. For $\rho_{ij} =0$,  $I_{ij}$ is zero, so we are left with
\begin{align}
I_{ij}(R) &= -\int_{-\infty}^{\infty}\mathrm{d}^2\w
\int_{0}^{\rho_{ij}}\mathrm{d}u
\, e^{-\half(\w_i^2 +\w_j^2  + 2u\w_i\w_j)}\,, \\
&=
-2\pi\int_{0}^{\rho_{ij}}\mathrm{d}u
\frac{1}{\sqrt{1-u^2}}\,,\\
&= -2\pi\arcsin\rho_{ij}\,, \\
&= -2\pi\arcsin\frac{R_{ij}}{\sqrt{R_{ii}R_{jj}}}\,. 
\end{align}

\setcounter{equation}{0}
\section{Computation of $I_{ijkl}$}
$I_{ijkl}(R)$ is given by
\be
\frac{1}{\pi^4}I_{ijkl}(R) =\frac{1}{\pi^4}\int_{-\infty}^{\infty}\mathrm{d}^4\w
\frac{1}{\w_i \w_j \w_k \w_l}  \,
e^{-\half \sum\limits_{m,n}\w_m R_{mn} \w_n}\,.
\ee
We rescale the $\w$ variable so that the matrix $R$ has unit numbers in the diagonal
$\w_i = \frac{\sqrt{2} \w_i}{\sqrt{R_{ii}}}$ and use
\be \label{tau_integral}
\frac{1}{\w_i} = \frac{\w_i}{\w_i^2} =  \w_i\int_{0}^{\infty}\mathrm{d}\tau e^{-\tau \w_i^2}\,,  
\ee
which combined  with the already existing $\w_i^2$ in the exponential gives
\be
\frac{1}{\pi^4}I_{ijkl}(R) =\frac{1}{\pi^4}
\int_{1}^{\infty} \mathrm{d}\tau
\int_{-\infty}^{\infty}\mathrm{d}^4\w
\frac{\w_i}{\w_j \w_k \w_l}  \,
e^{-\tau\w_i^2 - \sum\limits_{m\neq i}\w_m^2 - 2\sum\limits_{m<n}\w_m \rho_{mn} \w_n}\,,
\ee
with  (no assumed summation of repeated indices)
\begin{align}
\rho_{mn} &= \frac{R_{mn}}{\sqrt{R_{mm}R_{nn}}}\,,\nonumber\\
\label{rho_matrix}&= \frac{\Delta_{mn}}{\sqrt{\Delta_{mm}\Delta_{nn}}}\,,
\end{align}
where $\Delta_{mn}$ is the $(m,n)$ cofactor of the matrix $\Sigma$. 

We set the functions
\begin{align}
f(\wi) &= \wi e^{-\tau \wi^2}\,, \nonumber \\
g(\wi) &= e^{-2 \wi \left( \rho_{ij}\wj + \rho_{ik}\wk + \rho_{il}\wl\right)}\,, \nonumber
\end{align}
and perform an integration by part whose first contribution is zero and the remaining integral is
\begin{align}
\frac{1}{\pi^4}I_{ijkl}(R) &=-\frac{1}{\pi^4}
\int_{1}^{\infty} \frac{\mathrm{d}\tau}{\tau}
\int_{-\infty}^{\infty}\mathrm{d}^4\w
\left( \frac{\rho_{ij}}{\w_k\w_l} + \frac{\rho_{ik}}{\w_j\w_l} + \frac{\rho_{il}}{\w_j\w_k} \right)
e^{-\tau\w_i^2 - \sum\limits_{m\neq i}\w_m^2 - 2\sum\limits_{m<n} \w_m \rho_{mn} \w_n}\,, \nonumber \\
&=-\frac{1}{\pi^4} 
\int_{1}^{\infty} \frac{\mathrm{d}\tau}{\tau}
\left( \rho_{ij} F_{kl}^{ij}(\tau) + \rho_{ik} F_{jl}^{ik}(\tau) 
+ \rho_{il} F_{jk}^{il}(\tau) \right)\,,\nonumber
\end{align}
 with
\be
F_{kl}^{ij}(\tau) = \int_{-\infty}^{\infty}\mathrm{d}^4\w
\frac{1}{\w_k \w_l}
e^{-\tau\w_i^2 - \sum\limits_{m\neq i}\w_m^2 - 2\sum\limits_{m<n}\w_m \rho_{mn} \w_n}\,.
\ee
We integrate first the $\w$ variables that don't appear in the denominator and get
\begin{align}
F_{kl}^{ij}(\tau) &= 
\int\limits_{-\infty}^{\infty}\mathrm{d}\w_l
\int\limits_{-\infty}^{\infty}\mathrm{d}\w_k
\frac{e^{-\w_k^2 -\w_l^2 -2\rho_{kl}\w_{k} \w_{l} }}{\w_{k} \w_{l}}
\int\limits_{-\infty}^{\infty}\mathrm{d}\w_j
e^{-\w_j^2 -2\w_{j}\left(  \rho_{jk}\w_k + \rho_{jl}\w_l\right)}
\nonumber \\
&\int\limits_{-\infty}^{\infty}\mathrm{d}\w_i
e^{-\tau \w_i^2 -2\w_{i}\left(  \rho_{ij}\w_j +\rho_{ik}\w_k + \rho_{il}\w_l\right)}\,.
\end{align}
With
\be
\int_{-\infty}^{\infty}\mathrm{d}\w  e^{-a \w^2 + b w}
= \frac{\sqrt{\pi}}{\sqrt{a}}e^{\frac{b^2}{4a}}\,,
\ee
we obtain
\begin{align}
F_{kl}^{ij}(\tau) &= 
\int\limits_{-\infty}^{\infty}\mathrm{d}\w_l
\int\limits_{-\infty}^{\infty}\mathrm{d}\w_k
\frac{e^{-\w_k^2 -\w_l^2 -2\rho_{kl}\w_{k} \w_{l} }}{\w_{k} \w_{l}}
\int\limits_{-\infty}^{\infty}\mathrm{d}\w_j
e^{-\w_j^2 -2\w_{j}\left(  \rho_{jk}\w_k + \rho_{jl}\w_l\right)}
\frac{\sqrt{\pi}}{\sqrt{\tau}}e^{\frac{\left(\rho_{ij}\w_j +\rho_{ik}\w_k + \rho_{il}\w_l\right)^2}{\tau}} \nonumber \\
&=\frac{\sqrt{\pi}}{\sqrt{\tau}}
\int\limits_{-\infty}^{\infty}\mathrm{d}\w_l
\int\limits_{-\infty}^{\infty}\mathrm{d}\w_k
\frac{e^{-\w_k^2 -\w_l^2 -2\rho_{kl}\w_{k} \w_{l} + \frac{\left(\rho_{ik}\w_k + \rho_{il}\w_l\right)^2}{\tau} }}{\w_{k} \w_{l} }\,\nonumber \\
&\int\limits_{-\infty}^{\infty}\mathrm{d}\w_j
e^{-\w_j^2(1-\frac{\rho_{ij}^{2}}{\tau}) -2\frac{\w_{j}}{\tau}\left(  \w_k(\tau \rho_{jk} - \rho_{ij}\rho_{ik}) + \w_l(\tau \rho_{jl} - \rho_{ij}\rho_{il})\right)}  \nonumber \\
&=\frac{\pi}{\sqrt{\tau-\rho_{ij}^{2}}}
\int\limits_{-\infty}^{\infty}\mathrm{d}\w_l
\int\limits_{-\infty}^{\infty}\mathrm{d}\w_k
\frac{e^{-\w_k^2 -\w_l^2 -2\rho_{kl}\w_{k} \w_{l} + \frac{\left(\rho_{ik}\w_k + \rho_{il}\w_l\right)^2}{\tau} }}{\w_{k} \w_{l} }e^{\frac{\left(\w_k(\tau \rho_{jk} - \rho_{ij}\rho_{ik}) + \w_l(\tau \rho_{jl} - \rho_{ij}\rho_{il})\right)^2}{\tau(\tau-\rho_{ij}^2)}} \nonumber \\
&=\frac{\pi}{\sqrt{\tau - \rho_{ij}^2}}
\int\limits_{-\infty}^{\infty}\mathrm{d}\w_l
\int\limits_{-\infty}^{\infty}\mathrm{d}\w_k
\frac{1}{\w_k \w_l} e^{-\frac{\alpha_k(\tau) \w_k^2 + \alpha_l(\tau) \w_l^2 + 2\alpha_{kl}(\tau)\w_k\w_l  }{\tau - \rho_{ij}^2}}\,, \nonumber \\
&=
\frac{-2\pi^2}{\sqrt{\tau - \rho_{ij}^2}} 
\arcsin{\frac{\alpha_{kl}(\tau)}{\sqrt{\alpha_{k}(\tau)\alpha_{l}(\tau)}}}\,,
\end{align}
with
\begin{align}
\alpha_{k}(\tau) &= \tau\left(1- \rho_{jk}^2\right) + 2\rho_{ij}\rho_{ik}\rho_{jk} - \rho_{ij}^2 - \rho_{ik}^2 \,, \nonumber \\
&= \alpha_{k}^{0} \tau + \alpha_{k}^{1}\,, \nonumber \\
\alpha_{kl} &= \tau \left(\rho_{kl}- \rho_{jk}\rho_{jl}\right)  + \rho_{ij}\rho_{ik}\rho_{jl} + \rho_{ij}\rho_{il}\rho_{jk} - \rho_{ik}\rho_{il} - \rho_{ij}^2\rho_{kl}\,, \nonumber \\
&= \alpha_{kl}^{0} \tau + \alpha_{kl}^{1}\,, \nonumber \\
\alpha_{l} &= \tau \left(1- \rho_{jl}^2\right)  + 2\rho_{ij}\rho_{il}\rho_{jl} - \rho_{ij}^2 - \rho_{il}^2) \,, \nonumber \\
&= \alpha_{l}^{0} \tau + \alpha_{l}^{1}\,. \nonumber
\end{align}
Bringing all together we get
\begin{align}
\frac{1}{\pi^4}I_{ijkl}(R) &=\frac{4}{\pi^2}
\int_{0}^{1} \mathrm{d}u 
\left( 
\frac{\rho_{ij}}{\sqrt{1-\rho_{ij}^2 u^2}} 
\arcsin{\frac{\alpha_{kl}^{0} + \alpha_{kl}^{1}u^2}{\sqrt{(\alpha_{k}^{0} + \alpha_{k}^{1}u^2)(\alpha_{l}^{0} + \alpha_{l}^{1}u^2)}}} \right. \nonumber \\
&+
\frac{\rho_{ik}}{\sqrt{1-\rho_{ik}^2 u^2}} 
\arcsin{\frac{\alpha_{jl}^{0} + \alpha_{jl}^{1}u^2}{\sqrt{(\alpha_{j}^{0} + \alpha_{j}^{1}u^2)(\alpha_{l}^{0} + \alpha_{l}^{1}u^2)}}} \nonumber \\
\label{I4_final}&+\left.
\frac{\rho_{il}}{\sqrt{1-\rho_{il}^2 u^2}} 
\arcsin{\frac{\alpha_{jk}^{0} + \alpha_{jk}^{1}u^2}{\sqrt{(\alpha_{j}^{0} + \alpha_{j}^{1}u^2)(\alpha_{k}^{0} + \alpha_{k}^{1}u^2)}}} \right)\,,
\end{align}
 where we have made the change of variable $\tau = 1/u^2$.



\bibliography{bibliography}

\providecommand{\href}[2]{#2}\begingroup\raggedright\begin{thebibliography}{10}

\bibitem{Benhaddou2016}
K.~Benhaddou, \emph{{A Probabilistic Angle on One Loop Scalar Integrals}},
  \href{http://dx.doi.org/10.1088/1751-8121/aa6779}{\emph{J. Phys.} {\bf A50}
  (2017) 225202}, [\href{http://arxiv.org/abs/1603.05204}{{\tt 1603.05204}}].

\bibitem{Davydychev1992}
A.~I. Davydychev, \emph{General results for massive n‐point feynman diagrams
  with different masses},
  \href{http://dx.doi.org/http://dx.doi.org/10.1063/1.529914}{\emph{Journal of
  Mathematical Physics} {\bf 33} (1992) 358--369}.

\bibitem{SuzukiSantosSchmidt2003}
A.~T. Suzuki, E.~S. Santos and A.~G.~M. Schmidt, \emph{One-loop n -point
  equivalence among negative-dimensional, mellin–barnes and feynman
  parametrization approaches to feynman integrals}, {\emph{Journal of Physics
  A: Mathematical and General} {\bf 36} (2003) 11859}.

\bibitem{Fleischer2003}
J.~Fleischer, F.~Jegerlehner and O.~Tarasov, \emph{A new hypergeometric
  representation of one-loop scalar integrals in d dimensions},
  \href{http://dx.doi.org/http://dx.doi.org/10.1016/j.nuclphysb.2003.09.004}{\emph{Nuclear
  Physics B} {\bf 672} (2003) 303 -- 328}.

\bibitem{THOOFT1979}
G.~'t~Hooft and M.~Veltman, \emph{Scalar one-loop integrals},
  \href{http://dx.doi.org/http://dx.doi.org/10.1016/0550-3213(79)90605-9}{\emph{Nuclear
  Physics B} {\bf 153} (1979) 365 -- 401}.

\bibitem{vanOldenborgh1990}
G.~J. van Oldenborgh and J.~A.~M. Vermaseren, \emph{New algorithms for one-loop
  integrals}, \href{http://dx.doi.org/10.1007/BF01621031}{\emph{Zeitschrift
  f{\"u}r Physik C Particles and Fields} {\bf 46} (1990) 425--437}.

\bibitem{DENNER1991}
A.~Denner, U.~Nierste and R.~Scharf, \emph{A compact expression for the scalar
  one-loop four-point function},
  \href{http://dx.doi.org/http://dx.doi.org/10.1016/0550-3213(91)90011-L}{\emph{Nuclear
  Physics B} {\bf 367} (1991) 637 -- 656}.

\bibitem{VANOLDENBORGH1992}
G.~van Oldenborgh, \emph{The complex four point function for arbitrary masses},
  \href{http://dx.doi.org/http://dx.doi.org/10.1016/0370-2693(92)90499-T}{\emph{Physics
  Letters B} {\bf 282} (1992) 185 -- 189}.

\bibitem{DennerDittmaier2011}
A.~Denner and S.~Dittmaier, \emph{Scalar one-loop 4-point integrals},
  \href{http://dx.doi.org/http://dx.doi.org/10.1016/j.nuclphysb.2010.11.002}{\emph{Nuclear
  Physics B} {\bf 844} (2011) 199 -- 242}.

\bibitem{Phan2017ftp}
K.~H. Phan, \emph{{Scalar one-loop four-point Feynman integrals with complex
  internal masses}}, \href{http://dx.doi.org/10.1093/ptep/ptx079}{\emph{PTEP}
  {\bf 2017} (2017) 063B06}, [\href{http://arxiv.org/abs/1705.03602}{{\tt
  1705.03602}}].

\bibitem{Phan2017}
K.~H. Phan and T.~N.~H. Pham, \emph{{Scalar one-loop Feynman integrals with
  complex internal masses revisited}},
  \href{http://arxiv.org/abs/1710.11358}{{\tt 1710.11358}}.

\bibitem{DavydychevDelbourgo1998}
A.~I. Davydychev and R.~Delbourgo, \emph{{A Geometrical angle on Feynman
  integrals}}, \href{http://dx.doi.org/10.1063/1.532513}{\emph{J. Math. Phys.}
  {\bf 39} (1998) 4299--4334}, [\href{http://arxiv.org/abs/hep-th/9709216}{{\tt
  hep-th/9709216}}].

\bibitem{Davydychev2017}
A.~I. Davydychev, \emph{{Four-point function in general kinematics through
  geometrical splitting and reduction}},
  \href{http://arxiv.org/abs/1711.07351}{{\tt 1711.07351}}.

\bibitem{NISTDLMF}
``{\it NIST Digital Library of Mathematical Functions}.''
  http://dlmf.nist.gov/, Release 1.0.16 of 2017-09-18.

\bibitem{Childs1967}
D.~R. Childs, \emph{Reduction of the multivariate normal integral to
  characteristic form}, {\emph{Biometrika} {\bf 54} (1967) pp. 293--300}.

\bibitem{Hansen2014}
E.~W. Hansen, \emph{Fourier Transforms: Principles and Applications}.
\newblock Wiley, 1~ed., 2014.

\bibitem{Tilman2015}
T.~Butz, \emph{Fourier Transformation for Pedestrians}.
\newblock Undergraduate Lecture Notes in Physics. Springer International
  Publishing, 2~ed., 2015.

\bibitem{Davydychev2006}
A.~I. Davydychev, \emph{Geometrical methods in loop calculations and the
  three-point function},
  \href{http://dx.doi.org/http://dx.doi.org/10.1016/j.nima.2005.11.174}{\emph{Nuclear
  Instruments and Methods in Physics Research Section A: Accelerators,
  Spectrometers, Detectors and Associated Equipment} {\bf 559} (2006) 293 --
  297}.

\end{thebibliography}\endgroup
\bibliographystyle{JHEP}

\end{document}